# Interface dominated biferroic $La_{0.6}Sr_{0.4}MnO_3/0.7Pb(Mg_{1/3}Nb_{2/3})O_3 - 0.3PbTiO_3$ epitaxial superlattices

Ayan Roy Chaudhuri, R. Ranjith, and S.B. Krupanidhi[♣]
*Materials Research Centre, Indian Institute of Science, Bangalore 560012, INDIA*

R.V.K Mangalam, and A. Sundaresan
*Chemistry and Physics of Materials Unit, Jawaharlal Nehru Centre for Advanced Scientific Research, Jakkur, Bangalore 560064, INDIA*

**Abstract:**

Superlattices composed of ferromagnetic $La_{0.6}Sr_{0.4}MnO_3$ and ferroelectric $0.7Pb(Mg_{1/3}Nb_{2/3})O_3 - 0.3(PbTiO_3)$ layers were fabricated on (100) $LaAlO_3$ substrates by pulsed laser deposition technique. Ferromagnetic and frequency independent ferroelectric hysteresis characteristics established the biferroic nature of the superlattices. Influence of magnetic field was observed in tuning the P-E characteristics of the superlattices. Similar effect was observed on application of a high DC electric field to the samples. The nature of the observed ferroelectric properties and their modulation by applied magnetic and electric fields were thus discussed in connection to existence of dielectric passive layers at the ferroelectric/ferromagnetic interface.

---

[♣] Corresponding author: S. B. Krupanidhi
E-mail: sbk@mrc.iisc.ernet.in, ayan@mrc.iisc.ernet.in
FAX: +9180 2360 7316

Recently there is an upsurge of the quest for materials possessing coexisting magnetic and electrical order parameters[1]. A subset of these magnetoelectric/biferroic (ME) materials known as multiferroics, can exhibit a coupling between the ferroelectric (FE) and ferromagnetic (FM) order parameters, and their mutual control either by magnetic or electric field. Multiferroic materials are potential candidates for novel device applications with parametric values and flexibility unachievable in either FE or FM materials[2]. Several approaches have been made in order to synthesize new materials possessing these properties, besides the existing few intrinsic materials. One of them is designing engineered multilayers (ML) and superlattices (SL) with two or more distinct compounds[3]. These heterostructures can exhibit induced ferroelectricity and/or ferromagnetism due to breaking of time reversal symmetry. There are many reports on perovskite based artificial multiferroic superlattices with different manganates, such as $Pr_{1-x}Ca_xMnO_3$, $La_{1-x}Ca_xMnO_3$, $La_{1-x}Sr_xMnO_3$ etc. as magnetic material and $BaTiO_3$, $Ba_{1-x}Sr_xTiO_3$ etc. as the ferroelectric material[4-6]. Most recently there was a theoretical prediction of magnetoelectric effect in Fe/ $BaTiO_3$ ML.[7] To the best of our knowledge, there is no systematic report on the biferroic nature of superlattices composed of relaxor FE $0.7Pb(Mg_{1/3}Nb_{2/3})O_3 – 0.3PbTiO_3$ (PMNPT) and FM $La_{0.6}Sr_{0.4}MnO_3$ (LSMO) till date. In the present study epitaxial SL composed of LSMO and PMNPT have been fabricated using pulsed laser ablation, and their ferromagnetic, ferroelectric and effect of magnetic fields on ferroelectric properties have been reported.

SL of PMNPT/LSMO were grown on $LaNiO_3$ (LNO) coated (100) oriented $LaAlO_3$ (LAO) using stoichiometric ceramic targets. SL with different periodicities ($l$ = 6nm, 9nm, 13nm, 16nm) were fabricated keeping the magnetic layer thickness constant (~2nm) and increasing the thicknesses of the ferroelectric sublayers. The total thickness of the SL was set to be 250 ± 15 nm. Details of the growth conditions were reported elsewhere[8]. The θ-2θ and Φ scans were recorded on the sample for crystallographic and epitaxial characterizations using Phillips X'Pert MRD Pro X Ray diffractometer (CuK$_\alpha$ λ = 0.15418 nm). The magnetization hysteresis (M-H) was measured using a vibrating sample magnetometer in a PPMS system (by Quantum design, USA). A Radiant Technology Precision ferroelectric workstation was used to measure the room temperature ferroelectric polarization hysteresis. Dielectric properties of the heterostructures were measured within a frequency range of $10^2$-$10^6$ Hz using an Agilent 4294A precision impedance analyzer in the temperature range of -120°C to 400°C. For all the electrical measurements current perpendicular to the plane geometry has been used with LNO bottom electrode and gold (Au) dots of area $1.96\times10^{-3}$ cm$^2$ as top electrode.

Fig. 1 shows a typical *θ-2θ* scan around the (100) peak (18-25° in 2θ) of a SL with *l*=16 nm. Presence of higher order satellite peaks adjacent to the main peak, arising from chemical modulation of SL structure indicated that the films were heterostructurally coherent with sharp interfaces. The in-plane coherence of the SL was investigated by recording a Phi scan around the (103) plane of the substrate and the SL (inset of Fig. 1). Four distinct peaks with 90° spacing clearly indicate the fourfold symmetry as expected for cubic perovskite structures. This nature of the phi scans confirms the "cube on cube" epitaxial growth of SL. Broadness of the diffraction peaks in the SL phi scan shows that the films are strongly strained. Both LSMO (a = 3.87Å) and PMNPT (4.025Å) having respective lattice mismatches of -0.28% and -4.30% with the LNO (a= 3.859Å) experience compressive in plane stress. LSMO experiences also a tensile stress due to 3.73 % lattice mismatch with PMNPT.

The out of plane d spacing ($d_{100}$) of SL ranging between 4.08 and 4.19Å, gives an indication of the strain in the SL structures.

M-H loop measurements were performed on all the SL in order to characterize the ferromagnetic nature. Fig. 2 shows the M-H loop of a SL (*l*=16nm) measured at three different temperatures. The curves clearly exhibited well-defined coercivity confirming the ferromagnetic behaviour of the SL in the measured temperature range. Fig. 3(a) shows the room temperature polarization hysteresis (P-E) behaviour of the SL at probing frequencies ranging between 200 Hz to 2 kHz. Frequency independent and saturated natures of the P-E loops indicate the intrinsic ferroelectric nature of the samples. The P-E loops were asymmetric in nature and shifted towards the positive field axis. The shift was analyzed further by room temperature capacitance-voltage (*C-V*) measurements on the SL at 1 kHz frequency (inset of Fig. 3(a)). A similar shift in the C-V curves was observed along the voltage axis. *C-V* measurements performed before and after electrode annealing did not show any substantial cancellation of the voltage shift. The normalized asymmetry[9], L=$V_f/2V_c$ was calculated for the SL, where $V_f$ is the crossover voltage of the maximum capacitance value between the forward and reverse bias cycles and $2V_c$ is the difference between two voltages of maximum capacitance between rising and falling bias cycles in the *C-V* characteristics. With increasing SL periodicity, L decreased from 1.214 (d = 6 nm) to 0.455 (d = 16 nm), which could be attributed to the lattice strain relaxation. The P-E hysteresis loops were observed to be tilted along the field axis and the tilt decreases with increasing SL periodicity, accompanied by an increase in remnant polarization. These observations could be attributed to the presence of passive dielectric layers at the ferroelectric PMN-PT and ferromagnetic metallic LSMO interfaces[10]. The asymmetry of P-E loops has been discussed both theoretically and experimentally in connection to the passive layer formation at the ferroelectric/bottom electrode interfaces, and it was observed that ferroelectric asymmetry, which is a function of depolarizing fields and misfit dislocations, has a strong correlation to lattice relaxation at the nanoscale in heteroepitaxial thin films[10-13]. In the present case $d_{100}$ values decreased from 4.19 Å to 4.08 Å with increasing $l$ from 6nm to 16nm, which establishes the strain relaxation. The strain relaxation could be responsible for the observed asymmetry in the ferroelectric properties. Besides, asymmetric top and bottom electrodes can also be responsible for this kind of voltage shift[14]. In our case Au and LNO have been used as top and bottom electrodes. Room temperature dc leakage current (*I-V*) measurements, performed with Au as positive and negative electrodes alternatively were found to be symmetric in nature. This indicates that the shift is not primarily due to the asymmetric electrode configuration. In order to study the electrode effect on the polarization behaviour, it is essential to measure FE properties with different electrode materials. Systematic study of electrode effect on the FE properties of the PMNPT/LSMO superlattices is currently under progress.

Coexistence of room temperature ferromagnetic and ferroelectric properties in the heterostructures essentially proves the biferroic nature of the SL. To investigate the effect of an applied magnetic field on the ferroelectric properties, the biferroic superlattices were subjected to a magnetic field of 1.2T at room temperature with growth direction parallel to the applied magnetic field. Fig. 3(b) shows the room temperature P-E loops measured at a probing frequency of 1 kHz on the same top electrode of a superlattice before and after exposure to the magnetic field. The P-E loop measured after exposing the film to the magnetic field showed reduced asymmetry by shifting along the field axis and an increase in the remnant polarization and coercive field. The field axis shift defined as the difference between the coercive fields ($\Delta E_c$), before and after application of the magnetic field were 80.03 kV/cm (+$\Delta E_c$) and 96.68 kV/cm (-$\Delta E_c$). The observed effect could either be due to effective coupling between the residual magnetic moment in the SL and the applied electric field or it may be purely due to the effect of interfaces between the metallic FM and insulating FE layers. There can exist clamped magnetic domains in the material, creating charge accumulation at the interfaces. The accumulated charges could reduce the effect of the depolarizing field thereby reducing the asymmetry of the measured P-E hysteresis loops. A similar voltage shift phenomenon has been observed in epitaxial $BaTiO_3$ capacitors and is attributed to the existence of a non-ferroelectric layer at the interface between the ferroelectric layer and the bottom electrode[12]. In such cases, the crystal structure might be asymmetrically deformed in this layer by relaxation of lattice misfit strain[15]. In order to satisfy the continuity of the electric flux at the interface, the ferroelectric layer in one state (up or down) of polarization is more favourable than the opposite state. On switching the polarization from the more favourable to less favorable state by application of an electric field, the interface turns to a state with discontinuity of polarization, thereby generating a strong depolarizing field. Accumulation of free charge carriers at the interfaces between the non-switching layers and the ferroelectric layers causes the hysteresis loop to shift along the voltage axis. Fig. 3(c) shows the P-E hysteresis loops measured on a SL without any DC bias and after applying a 15 V DC bias. There was a clear shift of the hysteresis loops along the voltage axis on application of the DC bias with +$\Delta E_c$ = 46.73 kV/cm and -$\Delta E_c$ = 46.19 kV/cm. P-E measurement was repeated several times in absence of a DC bias in order to confirm that the shift observed after biasing is not an experimental artifact. The voltage axis shift increased with increasing DC bias. This establishes the presence of polarization discontinuity, which is compensated by injection of free charge carriers at the PMN-PT/LSMO interfaces. These experiments indicate in present case that the change in P-E hysteresis loops on application of a magnetic field is an interface dominated phenomenon.

To investigate the interface effect in further details, frequency dispersion behaviour of the SL structures has been studied. The nature of imaginary part of the dielectric constant ($\varepsilon''$) vs. frequency plots of the SL at different temperatures (Fig.4) shows $\varepsilon'' \to \infty$ as $\omega \to 0$. This exhibits a clear Maxwell Wagner (MW) type of relaxation for a superlattice according to which,

$$\varepsilon'' = \frac{1}{\omega C_0 (R_i + R_b)} \frac{1 - \omega^2 \tau_i \tau_b + \omega^2 \tau(\tau_i + \tau_b)}{1 + \omega^2 \tau^2}, (1)$$

where $\tau_i$, $\tau_b$, and $\tau$ are the relaxation times of interface, bulk PMNPT and the entire dielectric multilayer respectively. $C_0$ is a geometric factor, $R_i$ and $R_b$ are resistances of the interface and dielectric PMNPT and $\omega$ is frequency.

The observations indicate strong influence of interfaces on the observed dielectric and ferroelectric properties. The existence of MW type of relaxation in artificial superlattice structures has been studied extensively in ferroelectric BST SL and has been reported earlier both experimentally[16,17] and theoretically[18]. It has been observed that the presence of oxygen vacancy gives rise to a reduced resistivity of the interfacial material thereby effectively decreasing the resistivity of the whole SL structure[19]. The effective SL structure in the present case can be treated as a system with semiconductor interfaces intercalated between bulk insulating ferroelectric and conducting ferromagnetic layers.

The tuning of ferroelectric P-E loop on application of a magnetic field indicates the possibility of achieving substantial magnetoelectric coupling in these SL. Magnetocapacitance and magnetoloss characterizations with varying magnetic field and temperature are currently under progress in order to investigate the details of the ME behaviour of these superlattices.

In summary, epitaxial PMN-PT/LSMO superlattices were grown successfully on $LaNiO_3$ coated (100) oriented $LaAlO_3$ single crystal substrates by PLD technique. All the superlattice structures with different periodicities showed coexisting ferromagnetism and ferroelectricity at room temperature. The P-E hysteresis loops were asymmetric in nature. There was a distinct shift of the P-E loops along the voltage axis accompanied by an increase in remnant polarization and coercive field on application of a magnetic field. The shift of P-E loops along the voltage axis was also observed on application of DC bias to the samples. The asymmetric nature of the PE loops and their dependence on applied magnetic and dc electric fields indicated strong influence of the interfaces present in the heterostructures. The influence of the interfaces was further investigated by dielectric relaxation study in a wide range of temperature and frequency. The observed behaviour corresponds to the existence of Maxwell Wagner relaxation. All these observations in the present case suggest an interface dominated magnetically tuned ferroelectricity in the biferroic PMNPT/LSMO superlattices.

**Acknowledgement:**

The authors gratefully acknowledge Mr. Sandip Majumdar and Prof. Dr. S.K. Ray, department of physics and meteorology for experimental support and Prof. Dr. Wilfrid Prellier, Laboratoire CRISMAT CNRS UMR6508 ENSICAEN, France for his valuable suggestions.

**Figure Captions:**

Figure 1. X-Ray diffraction pattern of a PMN-PT/LSMO superlattice.

Inset: Φ scan of a LAO substrate and a superlattice.

Figure 2. M-H hysteresis loops of a PMN-PT/LSMO superlattice at three different temperatures.

Figure 3(a). Room temperature P-E hysteresis loops of PMN-PT/LSMO superlattice at different probing frequencies.

Inset: Room temperature C-V curve of the same superlattice.

Figure 3(b). P-E hysteresis loop of a superlattice before and after exposure to a magnetic field of 1.2T.

Figure 3(c). P-E hysteresis loop of a PMN-PT/LSMO superlattice before and after application of a dc bias.

Figure 4. $\varepsilon''$ vs. frequency characteristics of PMN-PT/LSMO superlattice at different temperatures.

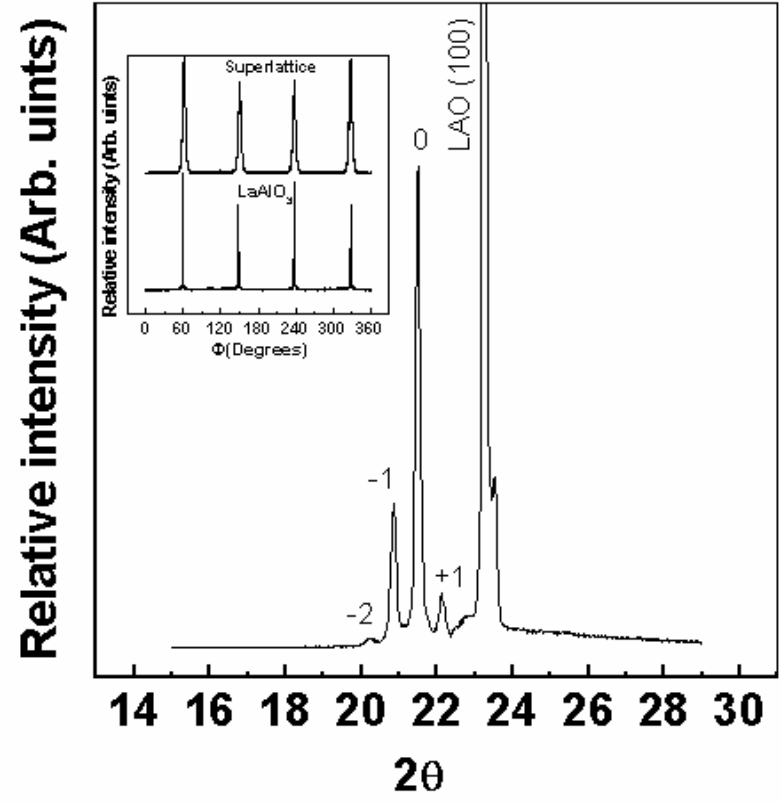

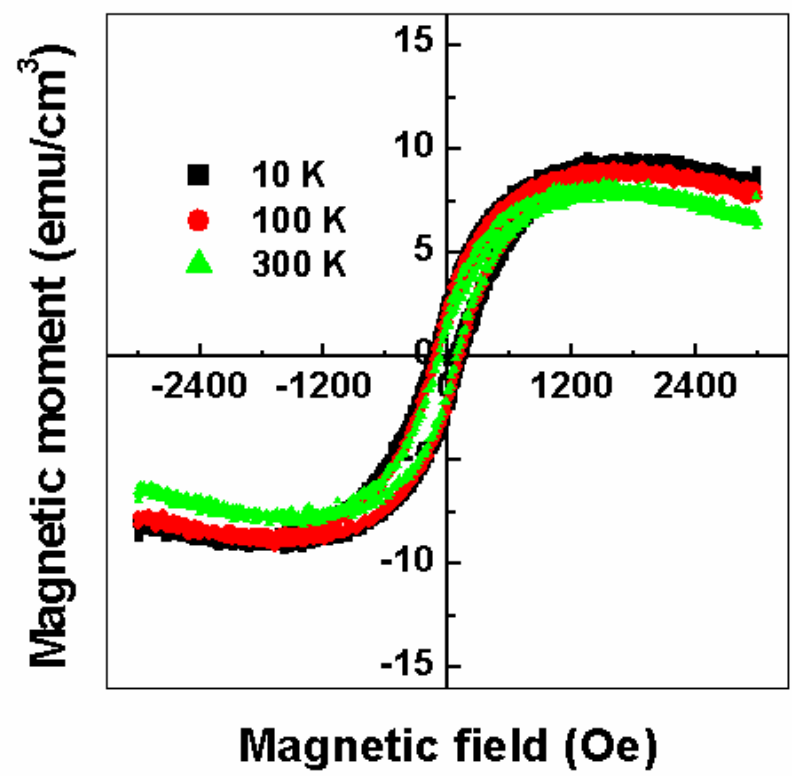

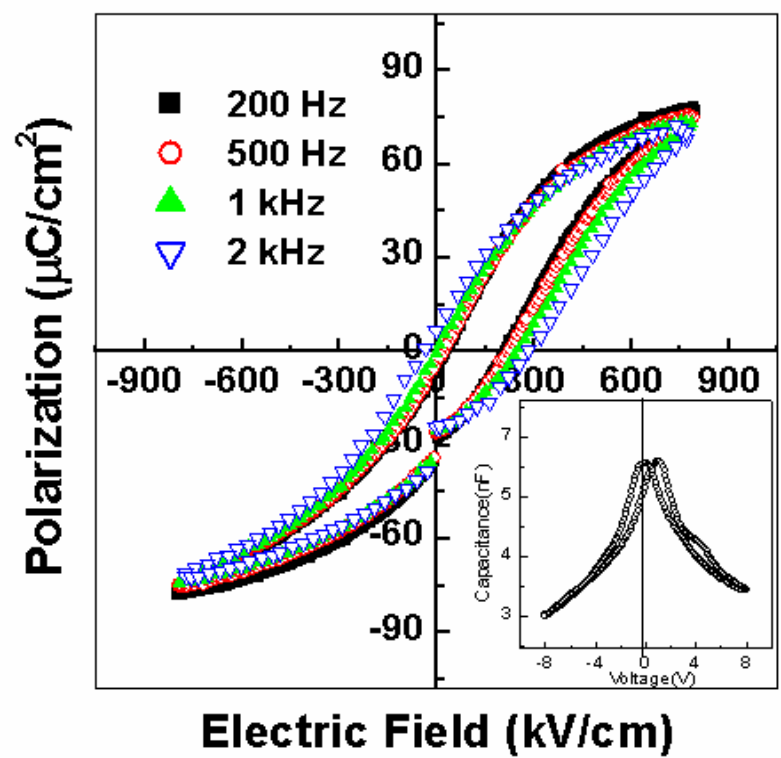

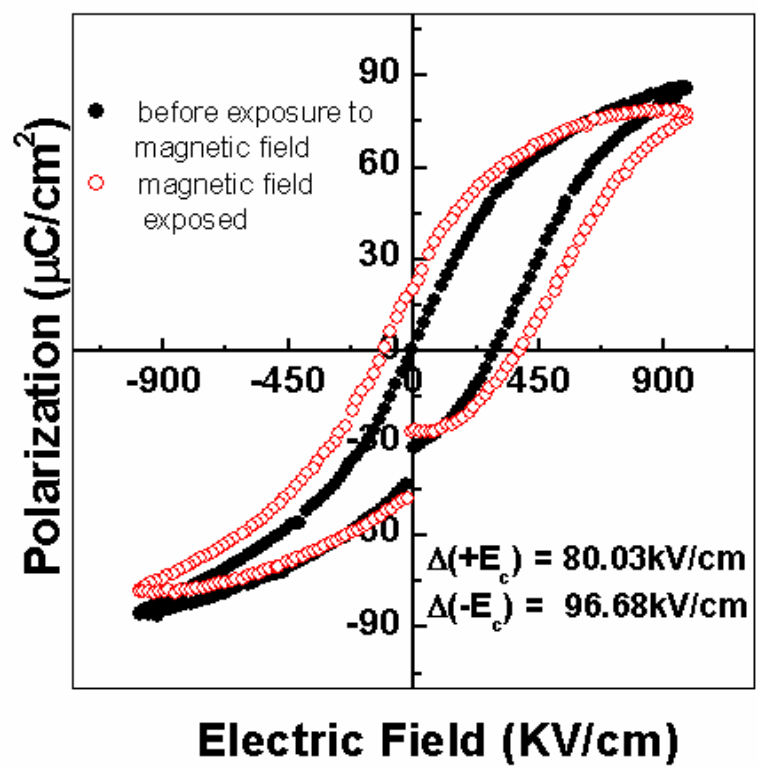

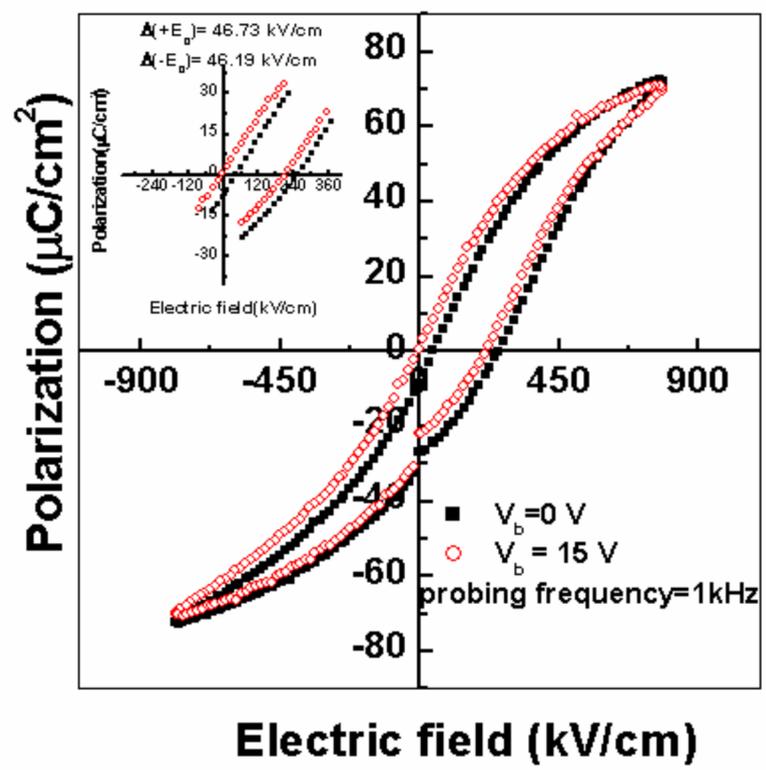

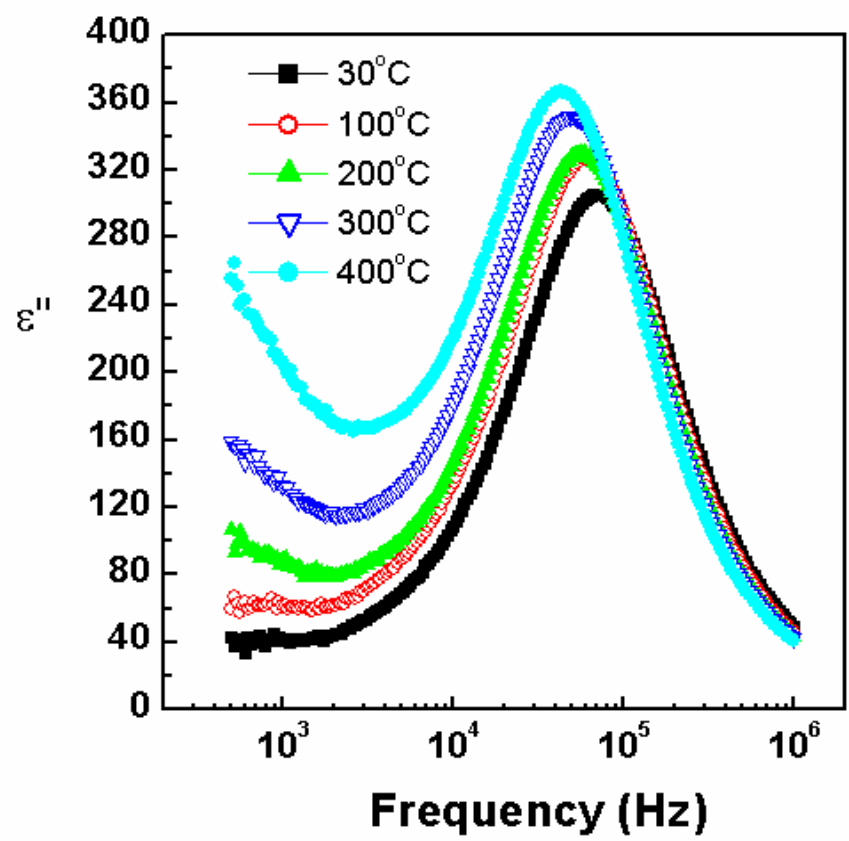